Malaria control and senescence: the importance of accounting for the pace and shape of aging in wild mosquitoes

Sadie J. Ryan[1,2,3,4,†], Tal Ben-Horin[5,6], Leah R. Johnson[7]

[1] Department of Geography, 3128 Turlington Hall, University of Florida, Gainesville, FL 32611 USA; [2] Emerging Pathogens Institute, P.O. Box 117315, University of Florida, Gainesville, FL, 32611 USA; [3] Center for Global Health and Translational Science, Department of Microbiology and Immunology, SUNY Upstate Medical University, Syracuse, NY 13210 USA; [4] School of Life Sciences, College of Agriculture, Engineering and Science, University of Kwa-Zulu Natal, Durban, 4000 South Africa; [5] Donald Bren School of Environmental Science & Management, University of California, Santa Barbara, California 93106 USA; [6] Haskin Shellfish Research Laboratory, Institute of Marine and Coastal Sciences, Rutgers University, 6959 Miller Ave, Port Norris, NJ 08349 USA; [7] Department of Integrative Biology, University of Southern Florida, Tampa, FL 33620 USA

†Email: sjryan@ufl.edu

**Abstract**

The assumption that vector mortality remains constant with age is used widely to assess malaria transmission risk and predict the public health consequences of vector control strategies. However, laboratory studies commonly demonstrate clear evidence of senescence, or a decrease in physiological function and increase in vector mortality rate with age. Despite recognition of its importance, practical limitations have stifled definitive observations of mosquito senescence in the wild, where rates of extrinsic mortality are much higher than those observed under protected laboratory conditions. We developed methods to integrate available field data to understand mortality in wild *Anopheles gambiae*, the most import vector of malaria in sub-Saharan Africa. We found evidence for an increase in rates of mortality with age. As expected, we also found that overall mortality is far greater in wild cohorts than commonly observed under protected laboratory conditions. The magnitude of senescence increases with *An. gambiae* lifespan, implying that most wild mosquitoes die long before cohorts can exhibit strong senescence. We reviewed available published mortality studies of *Anopheles spp.* to confirm this fundamental prediction of aging in wild populations. Senescence becomes most apparent in long-living mosquito cohorts; cohorts with low extrinsic mortality, such as protected laboratory cohorts, suffer a relatively high proportion of senescent deaths. Imprecision in estimates of vector mortality and changes in mortality with age will severely bias models of vector borne disease transmission risk, such as malaria, and the sensitivity of transmission to bias increases as the extrinsic incubation period of the parasite decreases. While we focus here on malaria, we caution that future transmission models of anti-vectorial interventions must incorporate both realistic mortality rates and age-dependent changes in mortality.

**Introduction**

Despite efforts to control the disease, the global malaria burden remains enormous. Each year there are over half a billion cases of malaria resulting in approximately one million deaths (Snow et al. 2005). An understanding of basic vector biology, when combined with advances in surveillance and treatment, offers the potential to provide rapid assessment of disease control for intervention (Hay et al. 2009, Mueller et al. 2011, White et al. 2011), particularly via vector control measures. A major challenge that remains is the accurate incorporation of mosquito life history into predictions of malaria risk and efficacy of control (Johnson et al. 2014). Life history variables, such as adult mortality, can often only be approximated, either from controlled laboratory experiments or field observations that often rely on unrealistic assumptions (Gillies and Wilkes 1965, Clements and Paterson 1981). The effects of adult mosquito mortality, for example, are realized at several stages in the pathogen transmission cycle, and minor imprecision can lead to substantial error in estimates of mosquito-borne pathogen transmission and malaria risk (Bellan 2010, Mordecai et al. 2013, Johnson et al. 2015). Since the majority of human malaria control programs are focused techniques that reduce adult mosquito survival (Dye 1992, Killeen et al. 2002), such as indoor residual spraying (IRS) with insecticides and the deployment of insecticide-treated nets (ITNs), error in estimates of mosquito disease risk can seriously undermine the efficacy of intervention.

Since the 1950s, the assumption that mosquitoes do not senesce, that is, that mortality and other traits remain constant with age, has been incorporated into models assessing the role of mosquitoes in pathogen transmission and to predict the public health consequences of vector control strategies. This assumption was first articulated by MacDonald (MacDonald 1957), who reasoned that extrinsic sources of mortality - such as predation and disease - kill mosquitoes long before they have an opportunity to die of old age. Constant mortality greatly simplifies the mathematics of models of mosquito-borne disease transmission and control. However, the rationale underlying this assumption opposes classical hypotheses of aging, asserting that senescence should proceed most rapidly when extrinsic mortality is great as it will select for shorter lifespan (Williams 1957, Baudisch 2005, Caswell 2007). Recent laboratory studies have challenged the constant mortality assumption, and demonstrate that factors such as environmental conditions, resource availability, and predation, influence how fast, and how much, mosquitoes senesce (Briegel and Kaiser 1973, Gary and Foster 2001, Afrane et al. 2006, Styer et al. 2007, Dawes et al. 2009). Studies of age-related changes in mosquito flight (Nayar and Sauerman Jr 1973), immune function (Christensen and Forton 1986), salivary gland structure (Beckett 1990), and detoxification mechanisms (Hazelton and Lang 1984, Lines and Nassor 1991), demonstrate that mosquitoes experience intrinsic deleterious functional and structural changes with age. Failure to account for senescence, in terms of mortality, in models of vector borne disease transmission will inflate estimates of disease risk (Styer et al. 2007), and overestimate the sensitivity of parasite transmission to interventions that reduce adult mosquito survival (Bellan 2010). Understanding adult mosquito mortality, including age-dependent changes in adult mortality rates, is critical given the current intervention paradigm of malaria reduction through vector control.

How mortality increases with age in wild mosquito populations is not well understood due to the difficulty of measuring mosquito mortality in nature. Standard mark-release-recapture (MRR) methods have been used most often to study mosquito mortality. However, this method has been challenged due to assumptions about equilibrium dynamics (Gaillard et al. 1994) and the movement of mosquitoes away from release sites (Harrington et al. 2001). Harrington et al. (2008) modified the standard MMR method, and tested the assumption of constant mortality by simultaneously releasing the yellow fever mosquito *Aedes aegypti* of different age cohorts and recapturing mosquitoes during a single event. This modified MRR method found evidence of increasing mortality with age, but was not designed to quantify the magnitude of senescence (Harrington et al. 2008). Methods to approximate physiological age, through gene expression (Hayes and Wall 1999), parity (Polovodova 1949, Detinova 1962), or cuticular hydrocarbon age grading (Hugo et al. 2006), can be used to assess senescence. Clements & Paterson (1981) analyzed existing physiological age distributions based on parity and found evidence of age-dependent mortality in wild populations of *Anopheles gambiae*, *An. funestus*, *An. arabiensis*, *An. farauti*, *An. vagus*, and *Culex quinquefasciatus*. However, simple approximations of physiological age prohibit direct comparisons of mortality in the wild to observations of mortality estimated from controlled laboratory studies. This is largely because age estimates based on parity assume the same rate of parity, or reproduction, and growth, in the lab and the wild. Given that the lab-reared mosquitoes are not resource restricted, this basic assumption is likely violated, and thus so is the proxy for chronological age.

Here we address patterns of mortality in wild *An. gambiae*, the most important vector of malaria in sub-Saharan Africa. We raise two questions: 1) is mortality in wild cohorts of *An. gambiae* age-dependent; and 2) how do patterns of mortality in the wild compare with mortality patterns observed under protected laboratory conditions? We address these questions in the context of vectorial capacity, a demographic index of a mosquito population's transmission potential, to quantify the impact of apparent mortality patterns on malaria risk. We developed methods to infer chronological age from physiologically age-structured data by modeling the link between physiological age and chronological age, and apply these methods to data presented by Gillies and Wilkes (1965) to estimate mortality in wild cohorts of *An. gambiae* in the vicinity of Muheza, Tanzania. We supplement this analysis by reanalyzing previously published *An. gambiae* mortality data from laboratory studies (Table 1), allowing direct comparisons of mortality in wild mosquitoes with laboratory observations of mosquito lifespan and the magnitude of mortality, as it increases with age.

**Materials and Methods**

*Age Distribution Model*

We developed methods to combine physiological age distributions sampled from wild mosquito populations, and MRR data determining the physiological age of mosquitoes of known chronological age. Unless otherwise noted, model simulations and analysis were conducted using MATLAB® VERSION 7.10 (The Mathworks, Natick, MA, USA).

First, we quantified the physiological age distribution of wild mosquitoes. Gillies and Wilkes (1965) surveyed the physiological age distribution of *An. gambiae* at Muheza, Tanzania over a period of fourteen months, from November 1962 to December 1963. Each month, indoor resting *An. gambiae* were caught and dissected, and physiological age was determined by the number of completed gonotrophic cycles. This is determined by examining the ovarian follicles from within the abdomen of dissected mosquitoes; when the ends of the tracheoles are tightly coiled, they are nulliparous, uncoiled, parous. Parity (gonotrophic cycle completion) is determined by examining the ovarioles for dilations, indicative of laying eggs. The maximum dilations counted per ovariole is taken as the number of completed gonotrophic cycles. While this research was conducted 50 years ago, this method is still in use today, refined and examined by multiple researchers (see details in Silver, 2007). We tested whether the physiological age distribution was constant across months in the study with pairwise multinomial exact tests implemented using the EMT package (Menzel 2010) in R VERSION 2.15.1. For each month, $i$, of the study period, we obtained the maximum likelihood multinomial proportions, $p_{0,i}, p_{1,i}, p_{2,i},…, p_{k,i}$, describing the proportion of total house-catches at each physiological age class, through the maximum physiological age class identified, $k$. We refer to these proportions as the projected physiological age distribution. We obtained 95% confidence intervals for the projected physiological age distribution using the GOFCI function (May and Johnson 2000) in R (R Core Team 2014).

Gillies and Wilkes (1965) extended their study of the *An. gambiae* population at Muheza to include MRR experiments determining the physiological age of recaptures of known chronological age. We took advantage of this data set to infer chronological age at each physiological age class. For each age class, we define chronological age to be a Poisson random variable, and found the maximum likelihood estimator (MLE) of the Poisson parameter and corresponding 95% confidence interval of the Poisson probability density function.

We assume that the projected physiological age distribution and the Poisson probabilities of chronological age for each physiological age class are independent. Then, by conditional probability, the product of realizations from these random variables can be used to generate realizations of the number of mosquitoes at each chronological age (i.e., the chronological age-distribution) via Monte Carlo simulation. We used a cohort size of 1000 mosquitoes for all realizations.

Mortality is reflected in the downward slope of the chronological age distribution, and the mortality hazard at age $a$, $H_a$, is indicated by the slope:

$$H_a = -\frac{d}{dt}\log(S_a) = -\frac{1}{S_a}\cdot\frac{dS_a}{dt} \quad (\text{eqn.1})$$

where $S_a$ is the number or proportion of mosquitoes at chronological age $a$. We calculated the mortality hazard from realizations of the chronological age distribution to estimate the instantaneous risk of mortality at chronological age.

The final step in our procedure is to determine whether or not the chronological age distribution exhibits patterns consistent with senescence. When the mortality rate of the population does not vary with age, a constant hazard function at age $a$ ($H_a$) can be written as:

$$H_a = \mu \quad (\text{eqn2})$$

where $\mu$ is the age-independent mortality rate. We refer to this age-independent hazard function as the constant mortality model. A variety of mortality patterns may emerge when the mortality rate increases with age. Kershaw et al (1954) observed that mortality in laboratory cohorts of mosquitoes is well described by the Gompertz mortality model, where the mortality hazard increases exponentially with age. The hazard function for the Gompertz mortality model can be written as:

$$H_a = \alpha e^{\beta a} \quad (\text{eqn3})$$

where α is the initial mortality rate and β is the exponential mortality increase with age. We use the Gompertz function here, because of the simple hazard function, although other monotonically increasing functions, such as logistic, are expected to yield qualitatively similar results.

We produced N = 10,000 realizations of the chronological age distribution and its associated mortality hazard. We found the MLEs for the parameters of both the constant and Gompertz mortality models for each realization, and then calculated the Akaike information criterion (AIC) (Burnham and Anderson 2002) for each realization to determine the best fit mortality model of for the observed mortality patterns. Taking the threshold of AIC ≥ 2 to denote meaningful differences, we interpret the model that is supported by the largest proportion of realizations to be the best overall model.

*The pace and shape of* **Anopheles gambiae** *mortality*

We seek to compare the observed mortality patterns in wild *An. gambiae* to patterns observed in laboratory cohorts and semi-wild cohorts either held in protected cages or marked and recaptured. We refer to wild cohorts held in protected cages as semi-field cohorts. We follow methods of Baudisch (2011) to compare patterns of mortality across laboratory, semi-field, and MRR studies, and examine two quantitative measures of mortality, the pace and shape of aging. These measures allow us to independently distinguish the rate of mortality and the magnitude of senescence. The pace of aging captures the time-scale on which mortality progresses, and is expressed by variables such as lifespan or longevity. The shape of aging does not depend on time; instead, it captures how sharply mortality increases with age, thereby independently quantifying the magnitude of senescence (Baudisch 2011). For the constant mortality model, the expected lifespan at adult emergence can be quantified as the inverse of the constant mortality rate:

$$L = \frac{1}{\mu} \quad (\text{eqn 4})$$

Expected lifespan for the Gompertz mortality model is defined as:

$$L = \frac{1}{\alpha} G\left(\frac{\alpha}{\beta}\right) \quad (\text{eqn 5})$$

where *G* is the exponential integral:

$$G(y) = y e^y \int_y^\infty \frac{1}{a} e^{-a} \text{ (eqn 6)}$$

We follow Baudisch (2011) and define a second measure of the pace of aging, longevity ($\Lambda$), as the age when 99% of adults have died and 1% are still alive. Several indices can be devised to capture the shape of aging, or magnitude of senescence. Here we describe shape as the ratio of lifespan to longevity ($L/\Lambda$). As this increases, so mortality increases more steeply with age.

We conducted a literature survey using the ISI Web of Science to uncover published estimates of *An. gambiae* mortality. We searched for papers published between 1945 and 2012 with titles, key words, or abstracts with the following general search strings: [(Anopheles) and (*gambiae* or *arabiensis* or *bwamae* or *merus* or *melas* or *quadriannulatus*) and (survival or mortality)]. Manuscripts were obtained for articles containing observations of mortality derived from monitoring mosquito cohorts under laboratory or semi-field conditions, and mortality observations from MRR experiments. Although the ISI Web of Science does not include grey literature, we believe our reference base was representative of the data available on the pace and shape of mosquito mortality relevant to this study.

We used DAGRA (Blue Leaf Software, Hamilton, New Zealand) to extract data from figures when the data were not available from tables, and recorded the study type (laboratory, semi-field, or MRR) and sex of mosquito cohorts used in each study. We estimated the hazard function when daily cohort survival data were presented, and tested the constant mortality assumption by fitting the constant and Gompertz mortality models as in the earlier analysis. Model fit was again evaluated by AIC. We used the parameter estimates from the best-fit mortality model in each case to estimate $L$, $\Lambda$, $L/\Lambda$, and $C_i^*$.

*Age-structured model of individual vectorial capacity*

We use a modified version of the age-dependent model of vectorial capacity developed by Styer et al (2007) to quantify the epidemiological implications of mortality patterns observed in wild *An. gambiae* populations. The model is based on the Ross-Macdonald formula (1957) for the basic reproductive number ($R_0$) of vector-borne diseases:

$$R_0 = \frac{mv^2 bp^{EIP}}{-r\ln(p)} \quad (\text{eqn } 7)$$

where *m* is the ratio of mosquitoes to humans, *v* is the biting rate (human bites per day per mosquito), *b* is the infectiousness of infected mosquitoes (proportion of bites that cause an infection), *p* is the daily survival rate of mosquitoes, *EIP* is the extrinsic incubation period of the malaria parasite, *Plasmodium falciparum* (number of days between a mosquito's infection and when it can yield infectious bites), and *r* is the recovery rate of human infectious cases or the inverse of the duration of infectiousness. $R_0$ can be described as the number of secondary cases generated by an index case in an otherwise susceptible population (MacDonald 1957). The mosquito related components of $R_0$ can be described by vectorial capacity (*C*):

$$C = \frac{mv^2 bp^{EIP}}{\ln(p)} \quad (\text{eqn } 8)$$

This index describes the transmission potential of the mosquito population while avoiding the difficulty associated with estimating the duration of infectiousness in humans (Garrett-Jones 1964). Vectorial capacity describes transmission intensity, and can be specified as the number of secondary infections that will be generated by a population of mosquitoes that is exposed to a single infectious human for one day. Both $R_0$ and *C* describe the potential for transmission in a completely susceptible and disease-free population: therefore, extrapolation to endemic systems should be taken with appropriate caution (Roberts 2007). Because we are primarily interested in the influence of mortality patterns on the mosquito population's transmission potential, we consider the biting rate, *v*, infection probability, *b*, and mosquito density, *m*, to be constant scalars that do not interact with changes in the probability of survival. Thus, in our analysis we consider a scaled version of vectorial capacity that is solely comprised of the extrinsic incubation period and probability of survival:

$$C^* = \frac{p^{EIP}}{-\ln(p)} \text{ (eqn 9)}$$

Scaled vectorial capacity is defined as the expected number of infectious biting days delivered by a mosquito population where all mosquitoes are infected. Mosquito populations with greater survival hold the potential to deliver more infectious bites, both because they are more likely to survive through the *EIP* and because they will live longer once infectious. Longer *EIP*s lead to fewer biting days with all else held constant.

The discrete, age-dependent model of vectorial capacity developed by Styer et al (2007) is given by

$$C^* = \sum_{a=\sigma}^{\infty} \Omega_a e_{a+EIP} \prod_{i=1+1}^{a+EIP} p_i \text{ (eqn 10)}$$

where $\Omega_a$ is the fraction of mosquitoes that are of age *a*, σ is the age at which mosquitoes begin biting, $p_a$ is the daily survival probability of a mosquito of age *a* (1-Ha). The term $e_{a+EIP}$ can be defined as the life expectancy of a mosquito of age *a+EIP*, or more intuitively, as the expected number of infectious biting days lived by a mosquito infected at age *a* if it survives the *EIP*. This term is calculated by summing under the discretized survival curve from age *a+EIP* forward:

$$e_{a+EIP} = \sum_{j=a+EIP+1}^{\infty} \prod_{i=a+EIP}^{j} p_i \text{ (eqn 11)}$$

Equation (10) reduces to equation (9) when mortality remains constant with age ($p_a = p$) and mosquitoes begin biting at age $a = 1$.

We simplify equation (11) by assuming that the age of first infectious blood meal is one day ($a = 1$), yielding scaled individual vectorial capacity ($C_i^*$), or the expected number of infectious biting days lived by a mosquito infected one day following adult emergence. We then use $C_i^*$ to quantify the impact of apparent mortality patterns on malaria risk. We obtain age-independent and age-dependent daily survival probabilities ($p_a$) from the parameter estimates of the best fit mortality model fit to the simulated data (as described in the previous section), and varied the EIP using values of 10, 20, and 30 days.

**Results**

*Age distribution model*

The physiological age distribution did not differ between months (pairwise multinomial exact tests; in all instances $P \geq 0.23$; Supplemental Table 1). The age composition of mosquitoes sampled at Muheza was dominated by nulliparous mosquitoes, comprising 31-50% of the monthly age compositions (Figure 1A). Mosquitoes completing at least six ovipositions comprised only 1.40-4.22% of the age composition each month. The majority of nulliparous mosquitoes recaptured in the MRR experiment were two days old, or less. However, nulliparous mosquitoes were recaptured up to five days post emergence (Figure 1B). The oldest recaptured mosquito was found 34 days following emergence and completed ten ovipositions. The remaining recaptured mosquitoes completed six or fewer ovipositions. The proportion of surviving mosquitoes declined sharply with chronological age, and the mean proportion of mosquitoes surviving to 30 days ranged from 0.01 in April 1963 to 0.02 in September 1963 (Supplemental Figure 1).

### *The pace and shape of* **Anopheles gambiae** *mortality*

The Gompertz mortality model best fit (AIC difference >2) the data in 9,585 of the realizations of the chronological age distribution (Supplemental Table 2); in 24 cases the simulation failed to converge, making this a 96.1% rate. The constant mortality model predicted a mean (±SD) rate of age-independent mortality (μ) of 0.18 (±0.049). Under the Gompertz model, the initial mortality rate (α) and exponential mortality increase with age (β) were negatively correlated (Pearson's ρ = -0.90, $P < 0.001$). Mean α was 0.11 (±0.07) and mean β was 0.03 (±0.04).

The Gompertz mortality model predicted an expected lifespan ($L$) of wild *An. gambiae* at Muheza to be 8.79 (±1.90) days, and predicted longevity ($Λ$) was 36.55 (±7.28) days, leading to a mean $L/Λ$ of 0.23 (±0.06). The constant mortality model predicts an expected lifespan of 5.78 (±1.53) days.

Our literature search yielded twelve studies investigating mortality in *An. gambiae* (Table 1), providing 38 examples of *An. gambiae* cohort survival. Twenty-two examples came from laboratory studies, eleven from semi-field studies, and five were from MRR experiments in wild cohorts. Among all mortality observations, we found $L$ to vary from 1.80 – 43.21 days. Lifespan ranged from 3.03 to 20.00 days when mortality was estimated from MRR data. We found evidence of both constant and age-dependent mortality when survival curves were available, though these data most commonly fit the age-dependent Gompertz mortality model, i.e., in no case was the constant model alone supported. Estimates of $L$ range from 2.55 – 43.21 days, (Figure 2) and $\Lambda$ from 11.44 – 115.26 days. $L/\Lambda$ varied from 0.22 (the approximate value of this ratio when mortality remains constant with age) to 0.74.

*Individual vectorial capacity*

In figure 3, the relationship between EIP and $C_i$ is depicted for both the Gompertz and Constant mortality models. Individual vectorial capacity under the constant mortality model increases as a monotonic function of $L$ (Figure 2, dashed line). This is consistent with the results of Styer et al (2007) and Bellan (2010) that the assumption of constant mortality leads to overestimates of vectorial capacity, particularly when the EIP is high. Under the Gompertz mortality model however, $C_i^*$ (and $L$) is a function of both Gompertz parameters and thus does not increase as a simple function of $L$ (Figure 2). We found that bias in $C_i^*$ arising from the constant mortality assumption increased with EIP (Figure 2A-C). We did not observe an association between $L/\Lambda$ and $C_i^*$ (in all cases Pearson's $\rho \leq 0.45$, $P \geq 0.07$; Supplemental Figure 2A-C), suggesting the contribution of senescence to $C_i^*$ is negligible when controlling for lifespan. However, a significant, positive correlation between $L$ and $L/\Lambda$ (Pearson's $\rho = 0.59$, $P = 0.01$; Figure 4) implies that the strength of senescence increases with lifespan, that is, age-dependent increases in mortality become most apparent when mosquitoes are long-lived.

**Discussion**

We found that mortality in wild/semi-wild *An. gambiae* can increase with age, and more importantly, is often far greater in wild/semi-wild cohorts than observed in cohorts held in protected laboratory conditions (Figure 2). When incorporated into models of vector disease transmission, such that those employed for malaria, an assumption of constant mortality with age leads to overestimates of transmission risk and the sensitivity of transmission to interventions that reduce adult mosquito survival, as demonstrated by Styer et al (2007) and Bellan (2010). Beyond validating assumptions about the shape of mortality, we also demonstrated that imprecise estimates of wild mosquito mortality will bias predictions of malaria transmission risk (as measured by scaled vectorial capacity). However this bias could approach, and even exceed that due to assumptions of constant mortality, particularly at low EIPs (Figure 2A). These results highlight the importance of properly translating mosquito and parasite life history into models of transmission risk when predicting and evaluating the effectiveness of disease intervention programs.

Increases in mortality with age have been observed in wild plant (Roach 2001, Roach et al. 2009) and animal populations (Promislow 1991, McDonald et al. 1996, Nussey et al. 2008), however, such survival patterns are only rarely demonstrated convincingly in wild insects, where rates of extrinsic mortality are commonly high (Bonduriansky and Brassil 2002). This underlies the paradigm of constant mosquito mortality, and is used conveniently in models of mosquito-borne disease transmission and control. This paradigm has been challenged previously (Briegel and Kaiser 1973, Gary and Foster 2001, Okech et al. 2003, Impoinvil et al. 2004, Afrane et al. 2006, Styer et al. 2007, Harrington et al. 2008, Dawes et al. 2009), and our results empirically demonstrate evidence of senescence in wild cohorts of *An. gambiae*, further opposing this paradigm. Mosquitoes and the diseases they transmit are complex, dynamic systems, driven by a combination of intrinsic processes such as age-dependent physiological deterioration and extrinsic factors such as climate and resource availability. However, we find the integrated effects of age-dependent and age-independent mortality factors on wild populations of *An.*

*gambiae* to minimize mortality increases with age, and senescence becomes most apparent in protected mosquito cohorts raised under suitable laboratory conditions (Figure 4).

Senescence can arise by physiological decline, independent of extrinsic mortality, or an interaction between age and environment that is manifest as an increasing susceptibility of older individuals to extrinsic factors. Several comparative studies of wild and captive plant and animal populations have indicated three emergent patterns: (1) senescence is most commonly caused by intrinsic physiological decline, rather than increasing vulnerability to extrinsic causes of mortality (Gaillard et al. 1994, Linnen et al. 2001, Ricklefs 2008); (2) aging-related mortality is often catastrophic, with individuals maintaining high levels of condition until shortly before their demise (Coulson and Fairweather 2001, Reed et al. 2008, Ricklefs 2008); and (3) populations of longer-lived individuals suffer a higher proportion of aging-related mortality (Roach 2001, Ricklefs 2008, Baudisch 2011). High extrinsic mortality in wild *An. gambiae* populations leads to few individuals surviving long enough to exhibit intrinsic physiological decline. By inherently controlling sources of mortality, laboratory and semi-field studies grossly underestimate rates of mortality, and overestimate the magnitude of senescence in wild mosquito populations.

Vector control interventions such as ITNs and IRS, using chemical insecticide or fungal biopesticide sprays, are the primary options available to contain and prevent pathogen transmission, and can cause substantial decreases in malaria transmission when vectorial capacity is high (Guyatt and Snow 2002). A critical factor influencing vectorial capacity is adult mosquito survival through the extrinsic incubation period of the parasite (EIP), ranging from approximately 10 to more than 30 days for malaria, depending on temperature (Patz and Olson 2006, Paaijmans et al. 2009). Only a relatively small fraction of mosquitoes naturally live long enough to infect humans when the EIP is long. Overestimating mosquito lifespan, combined with failing to account for age-dependent changes in mortality, will inflate predictions of vectorial capacity and the effectiveness of intervention at high EIP (Figure 2C). As EIP decreases, a greater proportion of mosquitoes survive through the EIP and the importance of accounting for senescence decreases (Figure 2A). Accurately translating observations of mosquito mortality into estimates of lifespan therefore assume increased importance at low EIP. The mortality rate we estimated from the constant mortality model is approximately double the rate used in a number of notable studies modeling malaria transmission risk (Martens et al. 1999, Parham and Michael 2010). A more accurate description of mortality in wild populations of *An. gambiae* will improve estimates of malaria risk and the efficiency by which control measures such as ITNs and IRS are administered and evaluated.

The phenomenon of age-dependent mortality, including its interactions with mosquito lifespan, is not limited to *An. gambiae*, and we expect this to manifest in all arboviral disease vectors. Due to the short EIP of many important arboviral pathogens, such as *Flaviviruses* (dengue, chikingunya), disease risk, and the impact of vector control, is exquisitely sensitive to small changes in vector mortality. The comparative analysis of mortality patterns in wild and laboratory populations of arthropod vectors will be essential to prioritize effective strategies for intervention and determine the human health risk posed by vector-borne disease.

**Acknowledgments**

This work was conducted as a part of the Malaria and Climate Change Working Group at the National Center for Ecological Analysis and Synthesis (NCEAS), a center funded by NSF (Grant #EF-0553768), the University of California, and the State of California. We are grateful to E. De Moore, K.D. Lafferty, E. Mordecai, A. McNally, K.P. Paaijmans, S. Pawar, and T. Smith for the initial discussion surrounding this work, and to C. Linkletter for assistance with the analysis.

**Table 1:** Predicted lifespan from laboratory, semi-field, and mark-recapture-release (MRR) studies investigating mortality in *Anopheles gambiae*. We fit constant and Gompertz mortality models to data from all studies presenting primary survival data (first five in the table), and the model that best fit the data, as determined by AIC, is described by Shape. All studies without primary survival data assumed constant mortality with age (italicized in table), so we did not fit those data to the Gompertz model.

| Study Type | Shape | Lifespan | Author |
|---|---|---|---|
| Lab | Constant & Gompertz | 2.484 – 30.221 | Bayoh (*unpublished*) |
| Lab | Gompertz | 15.441 – 37.344 | (Briegel and Kaiser 1973) |
| Semi-Field | Gompertz | 22.818 – 43.217 | (Gary and Foster 2001) |
| Semi-Field | Constant & Gompertz | 2.979 – 8.092 | (Ricklefs 2008) |
| Semi-Field | Constant & Gompertz | 6.207 – 13.581 | (Impoinvil et al. 2004) |
| *Lab* | *Constant* | *4.555* | *(Takken et al. 1998)* |
| *Lab* | *Constant* | *1.800 – 24.400* | *(Gary and Foster 2004)* |
| *Semi-Field* | *Constant* | *5.713* | *(Chege and Beier 1990)* |
| *MRR* | *Constant* | *5.685* | *(Costantini et al. 1996)* |
| *MRR* | *Constant* | *6.800* | *(Gillies 1961)* |
| *MRR* | *Constant* | *20.000* | *(Midega et al. 2007)* |
| *MRR* | *Constant* | *3.031* | *(Toure et al. 1998)* |

**Figures**

**Figure 1:** (A) Physiological age composition of *An. gambiae* house-catches at Muheza from November 1962 to December 1963. (B) Frequency distribution of chronological age at the time of recapture (open circles) for all physiological age classes. Solid lines are the fitted maximum likelihood Poisson probability density function (PDF). Data in (A) and (B) are presented in (Gillies and Wilkes 1965).

**Figure. 2:** Lifespan (*L* -days) and scaled individual vectorial capacity ($C_i$) for the constant mortality model (dashed lines). The points illustrate *L* and scaled $C_i$ estimated from laboratory (blue squares) and semi-field (green circles) observations of age-dependent *An. gambiae* mortality calculated using the appropriate value of EIP (A:10, B:20, C:30 days). The star illustrates *L* and $C_i$ estimated from the age-distribution model.

**Figure 3:** The relationship between vectorial capacity ($C_i$), estimated under the Gompertz and Constant mortality models, and the extrinsic incubation period (EIP), demonstrating the dramatic impact of these assumptions with increasing EIP, from 10 to 30 days.

**Figure 4:** Lifespan (*L* - days) and the magnitude of senescence (*L/Λ*) estimated under the Gompertz mortality model. The filled points illustrate *L* and (*L/Λ*) estimated from laboratory (blue squares) and semi-field observations (green circles), and the star illustrates *L* and (*L/Λ*) estimated from the age-distribution model. The dashed line indicates the constant mortality ratio assumption at approximately 0.217.

**Figure 1.**

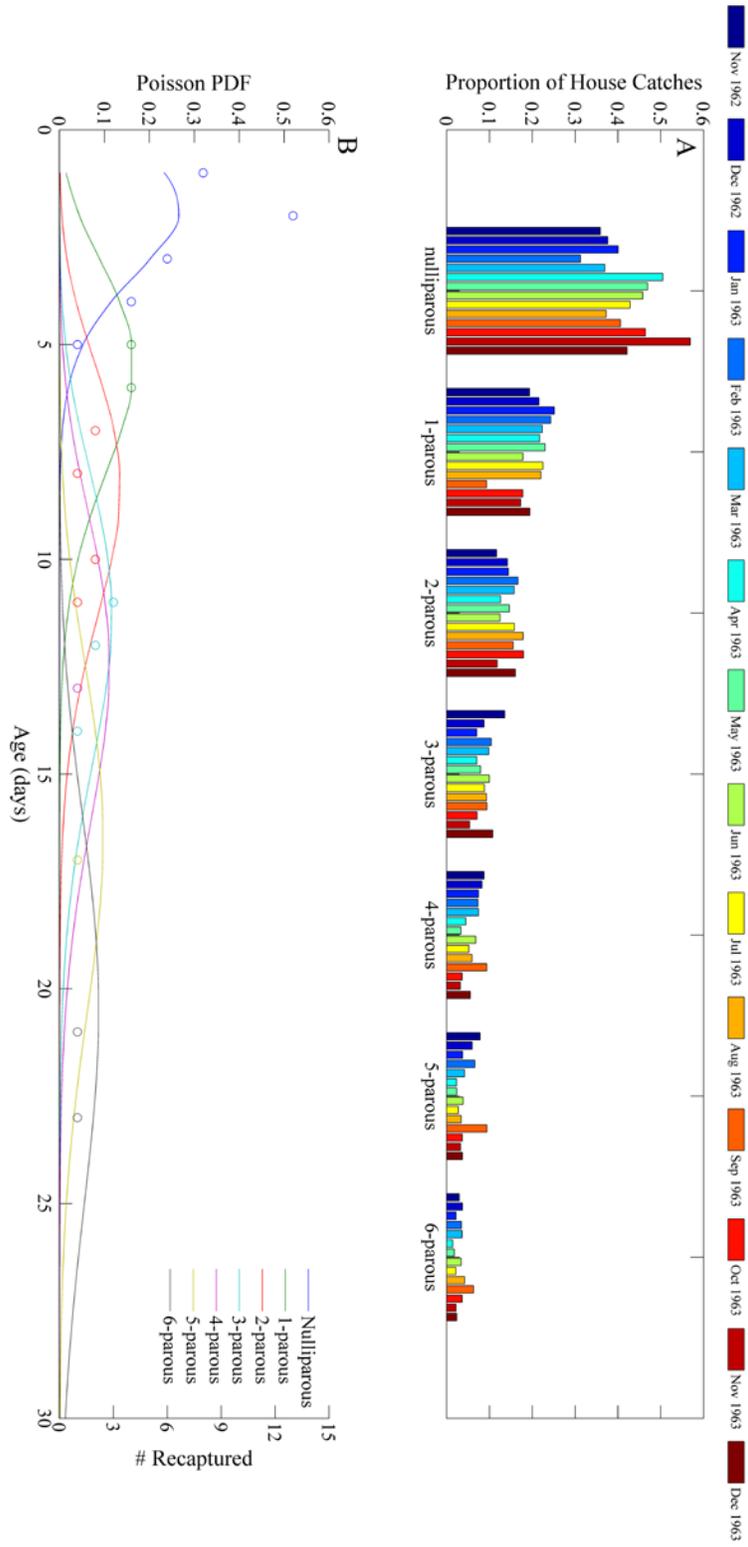

**Figure 2.**

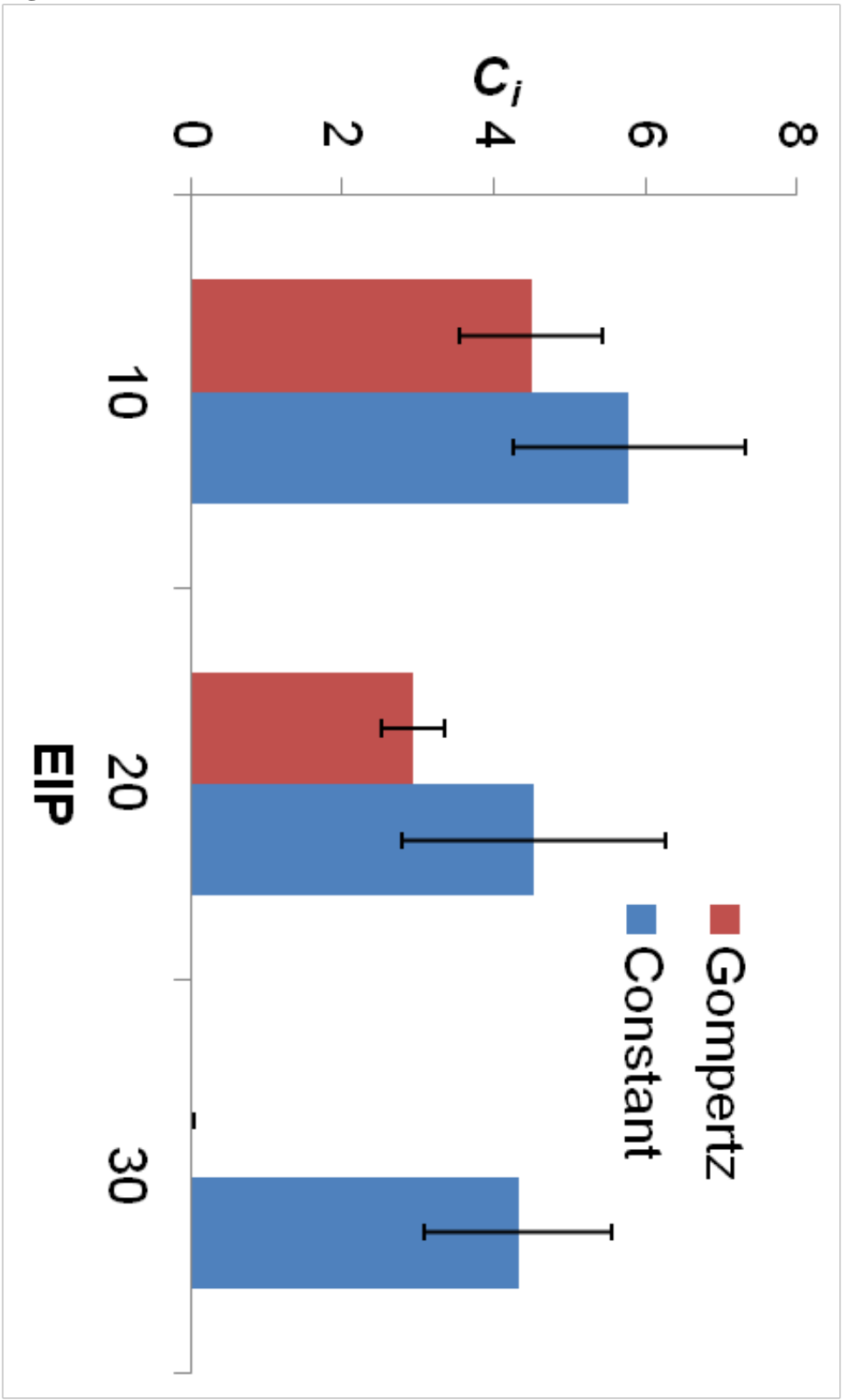

**Figure 3.**

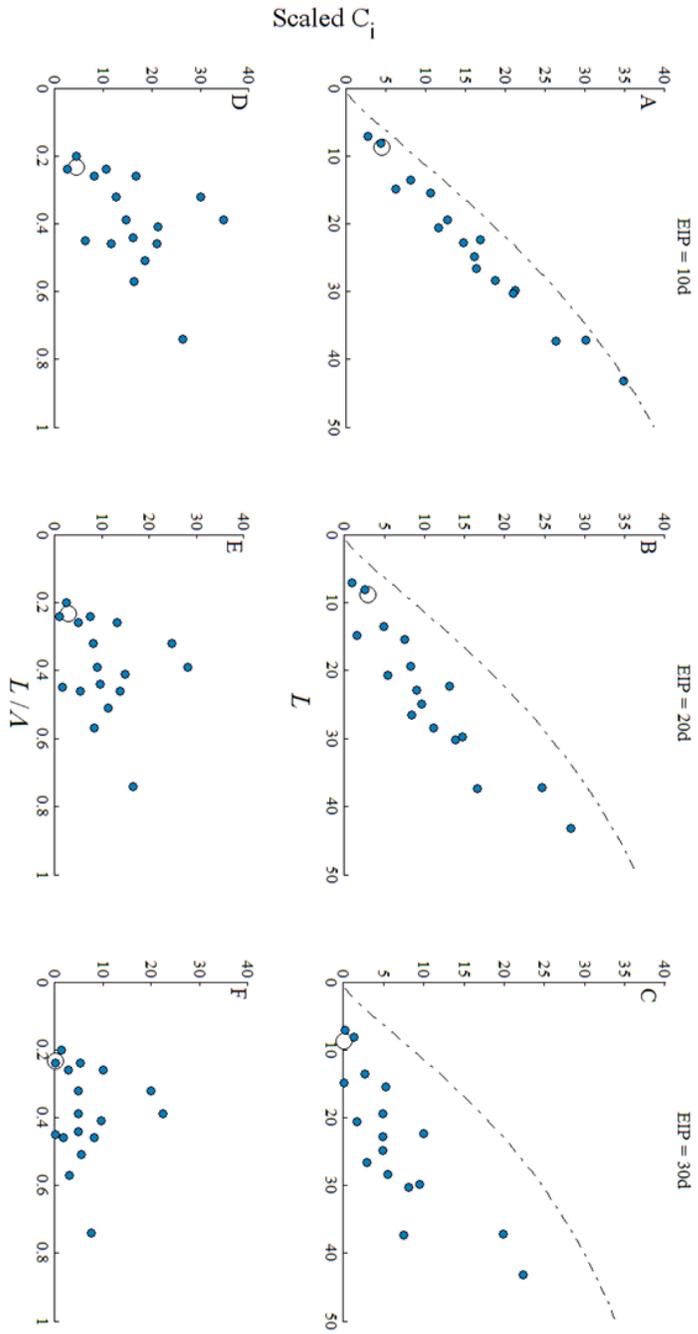

**Figure 4.**

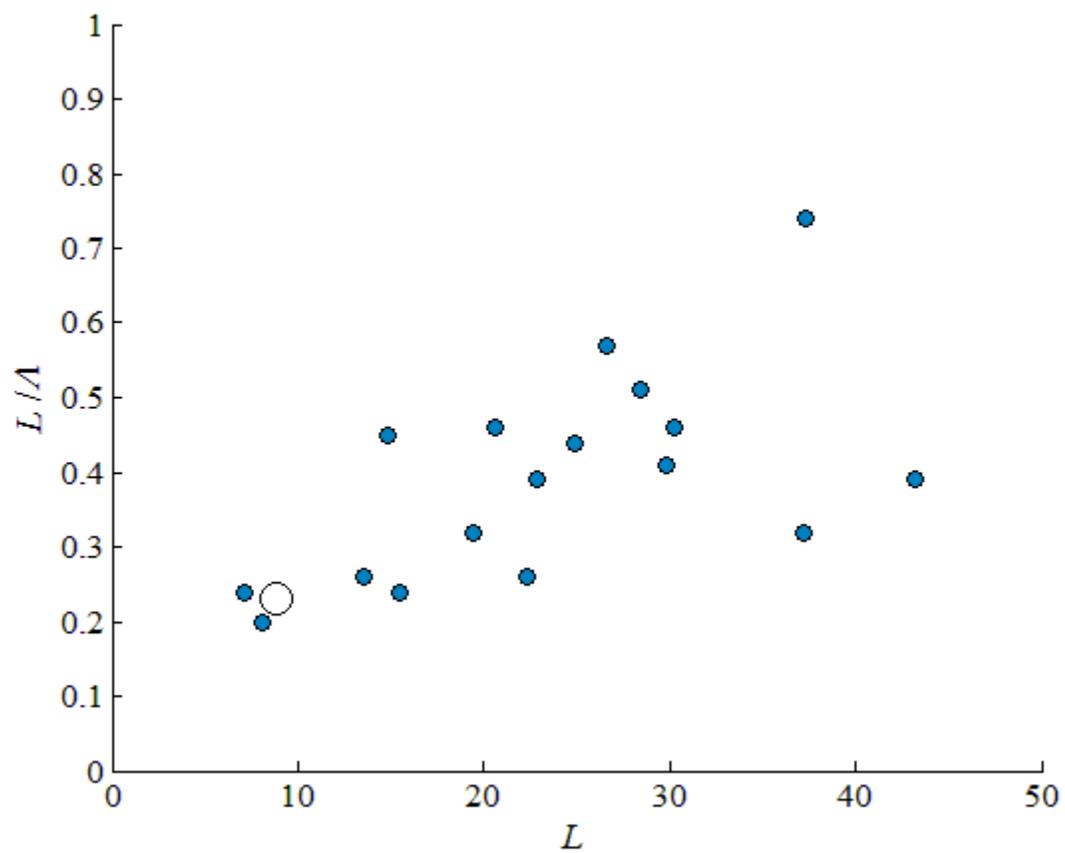

**Supplemental Figure 1**

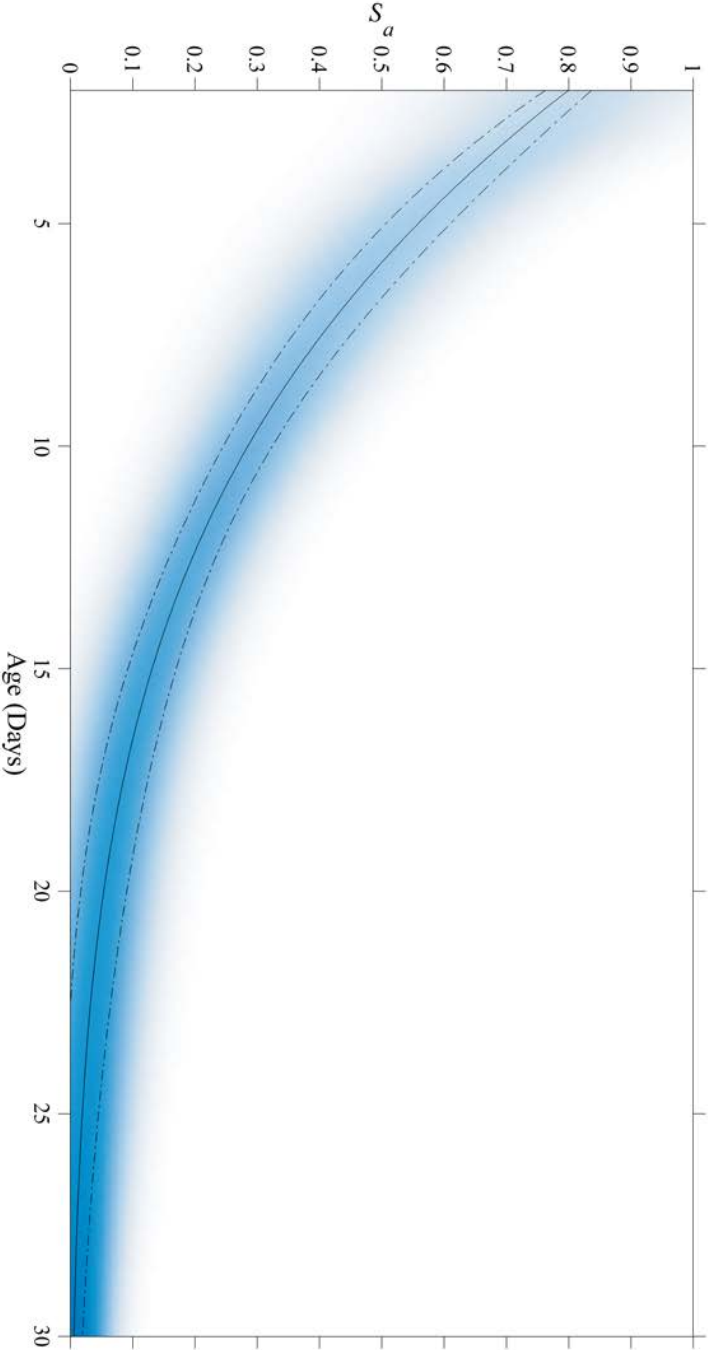

Supplemental Table 1: *P*-values of pairwise exact multinomial tests of differences in the physiological age distribution of *An. gambiae* recorded among monthly house-catches at Muheza, from November 1962 to December 1963

|  | Nov 1962 | Dec 1962 | Jan 1963 | Feb 1963 | Mar 1963 | Apr 1963 | May 1963 | Jun 1963 | Jul 1963 | Aug 1963 | Sept 1963 | Oct 1963 | Nov 1963 |
|---|---|---|---|---|---|---|---|---|---|---|---|---|---|
| Nov 1962 |  |  |  |  |  |  |  |  |  |  |  |  |  |
| Dec 1962 | 0.82 |  |  |  |  |  |  |  |  |  |  |  |  |
| Jan 1963 | 0.68 | 0.86 |  |  |  |  |  |  |  |  |  |  |  |
| Feb 1963 | 0.90 | 0.82 | 0.72 |  |  |  |  |  |  |  |  |  |  |
| Mar 1963 | 0.79 | 0.78 | 0.78 | 0.98 |  |  |  |  |  |  |  |  |  |
| Apr 1963 | 0.75 | 0.94 | 0.61 | 0.98 | 0.88 |  |  |  |  |  |  |  |  |
| May 1963 | 0.70 | 0.76 | 0.62 | 0.99 | 0.96 | 0.69 |  |  |  |  |  |  |  |
| Jun 1963 | 0.72 | 0.86 | 0.72 | 0.82 | 0.52 | 0.56 | 0.66 |  |  |  |  |  |  |
| Jul 1963 | 0.81 | 0.85 | 0.98 | 0.85 | 0.76 | 0.39 | 0.77 | 0.90 |  |  |  |  |  |
| Aug 1963 | 0.98 | 0.99 | 0.84 | 0.58 | 0.77 | 1.00 | 0.58 | 0.64 | 0.66 |  |  |  |  |
| Sept 1963 | 0.60 | 0.23 | 0.30 | 0.75 | 0.36 | 0.35 | 0.29 | 0.56 | 0.46 | 0.61 |  |  |  |
| Oct 1963 | 0.81 | 1.00 | 0.84 | 0.98 | 0.60 | 0.95 | 0.66 | 0.93 | 0.63 | 0.91 | 0.77 |  |  |
| Nov 1963 | 0.93 | 0.92 | 0.86 | 0.71 | 0.89 | 0.61 | 0.68 | 0.59 | 0.61 | 0.56 | 0.37 | 0.76 |  |
| Dec 1963 | 0.98 | 0.95 | 0.92 | 0.62 | 0.69 | 0.82 | 0.59 | 0.62 | 0.59 | 0.63 | 0.30 | 0.96 | 0.91 |